# Deep learning-based Fast Volumetric Image Generation for Image-guided Proton FLASH Radiotherapy

Chih-Wei Chang[1#], Yang Lei[1#], Tonghe Wang[2], Sibo Tian[1], Justin Roper[1], Liyong Lin[1], Jeffrey Bradley [1], Tian Liu[1], Jun Zhou[1] and Xiaofeng Yang[1*]

[1]Department of Radiation Oncology and Winship Cancer Institute, Emory University, Atlanta, GA 30322
[2]Department of Medical Physics, Memorial Sloan Kettering Cancer Center, New York, NY, 10065

[#]Co-first authors
[*]Corresponding email: xiaofeng.yang@emory.edu

**Running title:** Image-guided FLASH Radiotherapy
**Manuscript Type:** Original Research

**Abstract**

**Objective:** Proton FLASH therapy leverages ultra-high dose-rate radiation to enhance the sparing of organs at risk without compromising tumor control probability. This finding may allow us to push the limit-limiting constraint boundary and enable high prescription dose per treatment fraction. To prepare for the delivery of high doses to targets, we aim to develop a deep learning (DL)-based image-guide framework to enable fast volumetric image reconstruction for accurate target localization before FLSAH beam delivery.

**Approach:** The proposed framework comprises four modules, including orthogonal kV x-ray projection acquisition, DL-based volumetric image generation, image quality analyses, and water equivalent thickness (WET) evaluation. We investigated volumetric image reconstruction using four kV projection pairs with different source angles. Thirty lung patients were identified from the institutional database, and each patient contains a four-dimensional computed tomography (CT) dataset with ten respiratory phases. The leave-phase-out cross-validation was performed to investigate the DL model's robustness for each patient.

**Main results:** The retrospective patient study indicated that the proposed framework could reconstruct patients' volumetric anatomy, including tumors and organs at risk from orthogonal x-ray projections. Considering all evaluation metrics, the kV projections with source angles of 135° and 225° yielded the optimal volumetric images. The patient-averaged mean absolute error, peak signal-to-noise ratio, structural similarity index measure, and WET error were 75±22 HU, 19±3.7 dB, 0.938±0.044, and -1.3%±4.1%.

**Significance:** The proposed framework has been demonstrated to reconstruct volumetric images with accurate lesion locations from two orthogonal x-ray projections. The embedded WET module can be used to detect potential proton beam-specific patient anatomy variations. The framework can deliver fast volumetric image generation and can potentially guide treatment delivery systems for proton FLASH therapy.

# 1   Introduction

Proton therapy utilizes the physics characteristics of protons, which have well-defined ranges in a medium, to conformally deposit radiation energy to target volumes without exit doses (Lomax, 1999; Knopf and Lomax, 2013). This feature decreases the toxicity to healthy tissues such that patients who received proton treatment have lower unplanned hospitalization risks compared to photon treatment modalities (Baumann *et al.*, 2020). However, proton range uncertainty (Paganetti, 2012) requires additional margins for robust treatment planning (Albertini *et al.*, 2011; Zhou *et al.*, 2022a), which may compromise the sparing of healthy tissues when organs at risk (OAR) are consecutive to a treatment target region. In the era of precision medicine, the critical question is how to minimize the radiation doses to OAR such that the dose-limiting constraint no longer prevents the boost of prescription doses to the lesion volume.

Favaudon *et al.* (2014) demonstrated that ultra-high dose-rate ($\geq 40$ Gy/sec) radiation, the so-called FLASH effect, can increase tumor control probability (TCP) (Baumann and Petersen, 2005), while sparing normal tissues from acute radiation-induced apoptosis (Favaudon *et al.*, 2014). This promising finding can potentially make a paradigm shift in radiotherapy, and the FLASH effects have been widely explored (Esplen *et al.*, 2020; Wilson *et al.*, 2020; Gao *et al.*, 2022). Proton FLASH therapy has been investigated regarding the feasibility of using the current commercial treatment delivery system and inverse planning (Wei *et al.*, 2021; Kang *et al.*, 2022; Ma *et al.*, 2022). Meanwhile, the efficiency and accuracy of image-guided systems become important since proton FLASH treatment deposits high doses in target volumes (Diffenderfer *et al.*, 2020). The on-board fast imaging systems are essential to detect potential patient anatomy changes and motion management, especially for lung patients. However, the current proton on-board cone-beam computed tomography (CBCT) images require 30-60 seconds of scan time, and their quality is compromised due to motion and cavity artifacts.

Commercial proton machines, such as Varian ProBeam and IBA Proteus®ONE, include two kV x-ray sources with an image acquisition time of less than a second. The two orthogonal kV projections can potentially be acquired simultaneously, and the volumetric reconstruction method based on these projections will be free of motion and cavity artifacts. However, image reconstruction based on two projections is ill-conditioned. This ill-posed problem challenges the conventional mathematic image reconstruction methods. In contrast, Deep learning (DL) has been demonstrated as a universal approximator (Hornik *et al.*, 1989), and DL models feature in hierarchical learning to discover the underlying patterns behind the data (LeCun *et al.*, 2015). A significant challenge of applying DL to medical volumetric image reconstruction is the identification of tumor regions due to information lost when superimposing three-dimensional (3D) volumetric images to 2D projections (Yan *et al.*, 2016).

Many researchers have investigated various DL models to reconstruct 3D volumetric images based on limited 2D information (Li *et al.*, 2010; Shen *et al.*, 2019; Lei *et al.*, 2020). Another approach uses deformable image registration techniques to register 2D and 3D images (Zhao *et al.*, 2014). A recent interest development (Shao *et al.*, 2022) integrates DL and mechanical models to achieve real-time liver tumor localization. However, the previous literature is usually based on a single x-ray projection, and the robustness of DL models regarding CT numbers for dose evaluation remains an open question.

This study proposes a DL-based image-guide framework to inform the potential proton FLASH treatment, including tumor positions and patient anatomy changes. We use two orthogonal x-ray projections to provide additional information to enhance the predictability of DL models. Most importantly, we integrate a ray tracing-based water equivalent thickness (WET) evaluation module into the proposed framework for treatment feasibility investigation. This module can specifically detect potential anatomy changes corresponding to proton beams. To demonstrate the proposed framework, we investigate under which conditions the volumetric images can be derived effectively, accurately, and robustly to support medical decision-making.

## 2   Materials and methods

### 2.1   Patient data

This work aims to develop an image-guided framework to manage anatomy motion for accurate target localization before proton FLASH beam delivery. Since lung treatment usually requires motion management, we identify 30 patients from the institutional database with 4D CT for the framework demonstration. Each 4D CT dataset includes ten respiratory phases (CT) acquired from a Siemens SOMATOM Definition AS scanner using a 120-kVp spectrum. The CT dataset for each phase has a resolution of 0.98x0.98x3.0 mm$^3$. Synthetic kV x-ray projections were used to investigate the feasibility of the proposed framework, and the projections were generated based on Varian (Varian Medical Systems, Palo Alto) kV image system. The digital x-ray panel includes 768x1024 detector channels with a spacing of 0.39 mm. The synthetic kV projections were acquired from four different angle pairs, including 135°/225°, 157°/247°, 112°/202°, and 180°/270°, based on the x-ray source position.

### 2.2   Deep learning-based image-guided framework for proton FLASH treatment

The image-guided system is essential for proton FLASH treatment due to its ultra-high dose rate feature. Figure 1 depicts the DL-based image-guided framework for proton therapy FLASH treatment, including four modules to ensure the accuracy of target dose delivery. Figure 1(a) shows the kV image system of a conventional proton treatment machine, including two digital x-ray panels to acquire orthogonal projections. Figure 2(a) displays the volumetric image generation module using two orthogonal x-ray projections. A deep learning model, InverseNet3D, was implemented to demonstrate how to transform two orthogonal images into 3D images inversely. Section 2.2.1 gives the details of InverseNet3D regarding the model form and model parameters. Figure 3(c) shows the image evaluation module to quantify the integrity of generated volumetric images to conserve image features to reference CT, such as CT number histograms, image structures, noise levels, and statistical parameters. Section 2.2.2 gives the details of each evaluation metric. Figure 4 depicts the module for treatment evaluation to detect potential patient anatomy changes. The assessment is based on water equivalent thickness (WET) comparisons between the reference CT and generated volumetric images, and the details of comparisons are described in Section 2.2.3. The framework performance has been evaluated regarding image quality and proton characteristics (i.e., WET). The validated framework is expected to deliver on-board volumetric images to localize target locations to guide the delivery of proton FLASH treatment.

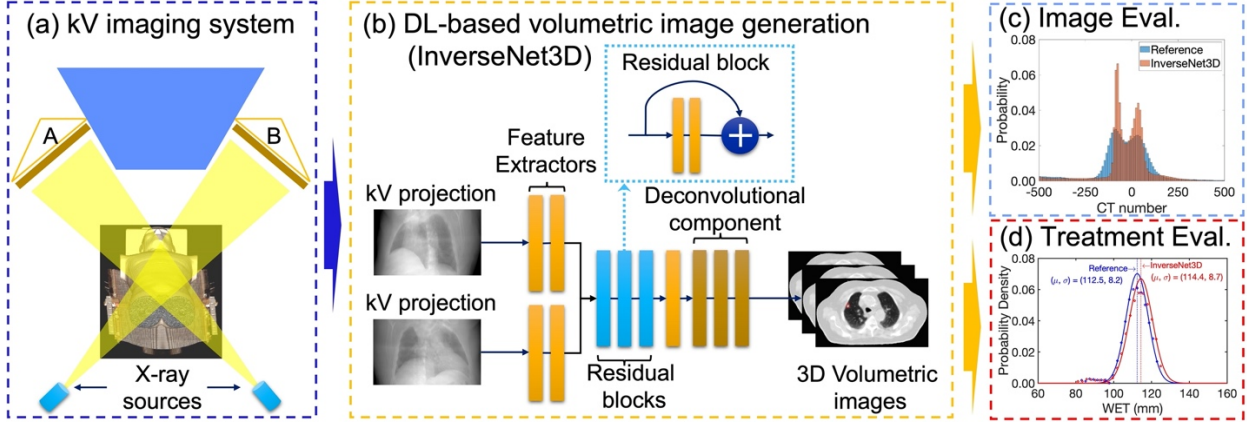

**Figure 1.** Deep learning-based image-guided framework for proton therapy FLASH treatment, including four modules: (a) kV image system with two digital x-ray panels A and B to acquire two orthogonal projections simultaneously, (b) Deep learning (DL)-based volumetric image generation using orthogonal kV projections (InverseNet3D is implemented in this module), (c) image evaluation based on CT numbers to ensure the integrity of generated volumetric images without systematic shift of voxel intensity, and (d) treatment evaluation based on water equivalent thickness (WET) to detector potential anatomy changes.

### 2.2.1 InverseNet3D

Figure 1(b) depicts the hierarchical structure of the InverseNet3D to transform two orthogonal x-ray projections inversely. InverseNet3D includes three components to infer volumetric images from 2D projections. Initially, the feature extractors, built by convolutional neural networks (CNN), are used to identify the local patterns from the x-ray projections. The second component includes multiple residual blocks to prevent gradient vanishing and enhance the performance of model learning during error backpropagation processes. Each residual block's fundamental unit is composed of two convolutional layers with a single residual layer. Ultimately, the deconvolution component upscales the dimension of local feature images received from the residual blocks to conserve the dimensions of volumetric images (CT) from patients. The detailed model structures and parameters are given in Appendix A.

To investigate the model robustness, the InverseNet3D was trained using leave-phase-out patient-specific training with 4D CT datasets from 30 patients. Supervised loss functions were used to train the model, including MAE loss for voxel-wised image intensity learning and gradient loss for image edge learning. Tensorflow v1.15.0 (Abadi *et al.*, 2016) was used to implement model hierarchy, optimization, and data preprocessing. The simulation environment included an NVIDIA Tesla V100 GPU.

### 2.2.2 Image evaluation

Three metrics are used to evaluate the quality of each volumetric image voxel generated by InverseNet3D within the region of interest (ROI). Eq. (1) gives the mean errors (ME) where the *x*, *i*, *N*, *DL*, and *ref* denote the voxel CT number, the $i^{th}$ voxel, total voxels, generated images by InverseNet3D, and reference images. The unit of CT numbers is Hounsfield units (HU). ME can check if the generated volumetric images have systematic intensity shifts from the reference. Eq. (2) defines the mean absolute errors (MAE) to determine the global quality of the generated images. Eq. (3) gives the peak signal-to-noise ratio (PSNR) to evaluate the reconstruction quality of the InverseNet3D from orthogonal 2D projections. The structural similarity index measure (SSIM), implemented from the literature (Zhou *et al.*, 2004), is used to quantify the structural consistency of lesion ROI between the reference and InverseNet3D. The histogram of CT numbers is also evaluated to investigate differences in global profiles. All evaluation metrics were implemented using MATLAB R2021a.

$$ME = \frac{1}{N}\sum_{i=1}^{N}(x_{i,DL} - x_{i,ref}) \tag{1}$$

$$MAE = \frac{1}{N}\sum_{i=1}^{N}|x_{i,DL} - x_{i,ref}| \tag{2}$$

$$PSNR = 10\log_{10}\left[\frac{\max^2(x_{ref})}{\frac{1}{N}\sum_{i=1}^{N}(x_{i,DL} - x_{i,ref})^2}\right] \tag{3}$$

### 2.2.3 Treatment evaluation

Due to the ultra-high dose rate feature of proton FLASH therapy, patient anatomy changes are critical and high impact the treatment quality. The water equivalent thickness (WET) (Zhang and Newhauser, 2009; Zhang et al., 2010; Newhauser and Zhang, 2015) can be used to quantify the potential anatomy changes during inter-fractional or intra-fractional proton FLASH treatment. To maximumly spare organs at risk, an anteroposterior beam is commonly used in the treatment (Wei et al., 2022), and we explore the WET for this beam. We implemented a ray tracing-based WET algorithm (Amanatides and Woo, 1987; Niepel et al., 2019) to derive the WET within the target ROI using MATLAB R2021a. The essential CT-number-to-relative-stopping-power (RSP) conversion table (Chang et al., 2020; Chang et al., 2022b) is given in Appendix B for WET calculation.

The RSP can be derived from CT numbers using HLUT in Appendix B. Gaussian fitting is used to fit raw WET histograms to minimize the uncertainty due to image noises. The WET uncertainty can cause by patient anatomy changes and inconsistent image acquisition modalities for treatment planning and daily image guidance. This work aims to investigate the WET uncertainty from an image-guided system (i.e., the quality of generated images from InverseNet3D). Eq. (4) defines the difference of WET (ΔWET) within a given ROI where $i$, $N$, $DL$, $ref$ denote the i[th] voxel in the ROI, total voxels in the ROI, InverseNet3D generated images, and reference images. Eq. (5) defines the relative difference of WET ($\varepsilon_{WET}$) within a given ROI. We focus on the target ROI in this work since the target volume is directly associated with proton beams. To increase tumor control probability, we want to ensure accurate dose delivery to the target volume.

$$\Delta WET = \frac{1}{N}\sum_{i=1}^{N}(WET_{i,DL} - WET_{i,ref}) \tag{4}$$

$$\varepsilon_{WET} = \frac{\Delta WET}{\frac{1}{N}\sum_{i=1}^{N} WET_{i,ref}} \times 100\% \tag{5}$$

## 3  Results

### 3.1  Volumetric image generation using orthogonal kV projections from different angles

The evaluation metrics for each patient are given in Table C1-C4 for volumetric image analyses by different orthogonal projection pairs. Table C1-C4 provide the phase-averaged outcomes for each patient. The results indicate that each metric's standard deviations (SD) are usually slight between raspatory phases. The SD values are generally smaller than 5% of the mean values.

Table 1 shows the patient average evaluation results of generated image quality using InversereNet3D with four orthogonal projection pairs. The 180°/270° projection pair results in the minimum ME, which is approximate -0.3% (-3.3/1000x100%) error regarding material properties because of *CT number (HU) = 1000(μ-1)*, where $\mu$ is the relative linear attenuation coefficient material to water. The percentage ME values for the other three orthogonal projection pairs are -1.6%, -0.9%, and 0.4%, corresponding to the angle pairs of 135°/225°, 157°/247°, and 112°/202°.

The generated volumetric image from the 112°/202° projection results in the optimal MAE, while the 180°/270° projection causes the maximum MAE. The PSNR analyses show that both 157°/247° and 180°/270° projections have comparable 3D reconstruction image quality. The volumetric image infers from the 112°/202° projection causes the minimal PSNR. The minimum SSIM achieves by the image generated by using InverseNet3D with the 157°/247° projections. The volumetric images inferred from the 180°/270° projection results in the worst SSIM value.

**Table 1.** Evaluation metrics of volumetric image quality generated from InverseNet3D using mean error (ME), mean absolute error (MAE), peak signal-to-noise ratio (PSNR), and structural similarity index measure (SSIM). All metrics are averaged over all patients given in Table C1-C4. ME and MAE are evaluated for the whole volume, and PSNR and SSIM are computed for the target contour.

|  | Orthogonal projections | | | |
| --- | --- | --- | --- | --- |
|  | 135° / 225° | 157° / 247° | 112° / 202° | 180° / 270° |
| ME (HU) | -15.5 ± 19.4 | -9.2 ± 15.3 | 3.9 ± 17.8 | -3.3 ± 15.0 |
| MAE (HU) | 75.6 ± 22.4 | 74.1 ± 21.3 | 73.4 ± 19.2 | 80.3 ± 24.0 |
| PSNR (dB) | 19.0 ± 3.7 | 20.2 ± 4.3 | 15.9 ± 3.6 | 20.4 ± 3.0 |
| SSIM | 0.938 ± 0.044 | 0.945 ± 0.046 | 0.941 ± 0.045 | 0.928 ± 0.058 |

Figure 2 demonstrates that InverseNet3D can generate the volumetric images from four orthogonal projection pairs using patient 3 in transversal, coronal, and sagittal views. The orange arrows indicate the lesion location for both reference and generated images. Volumetric images derived from all orthogonal projection pairs can reconstruct the patient's anatomy and identify the tumor in the lung. The heart and liver can be clearly identified from the transversal and sagittal images. Figure 3 depicts the reference and generated images for patient 15, which has a smaller body and tumor size than the patient shown in Figure 2. The lesion target and important organs can be recognized from the volumetric images, such as the tumor, heart, aorta, vena cava, and spine.

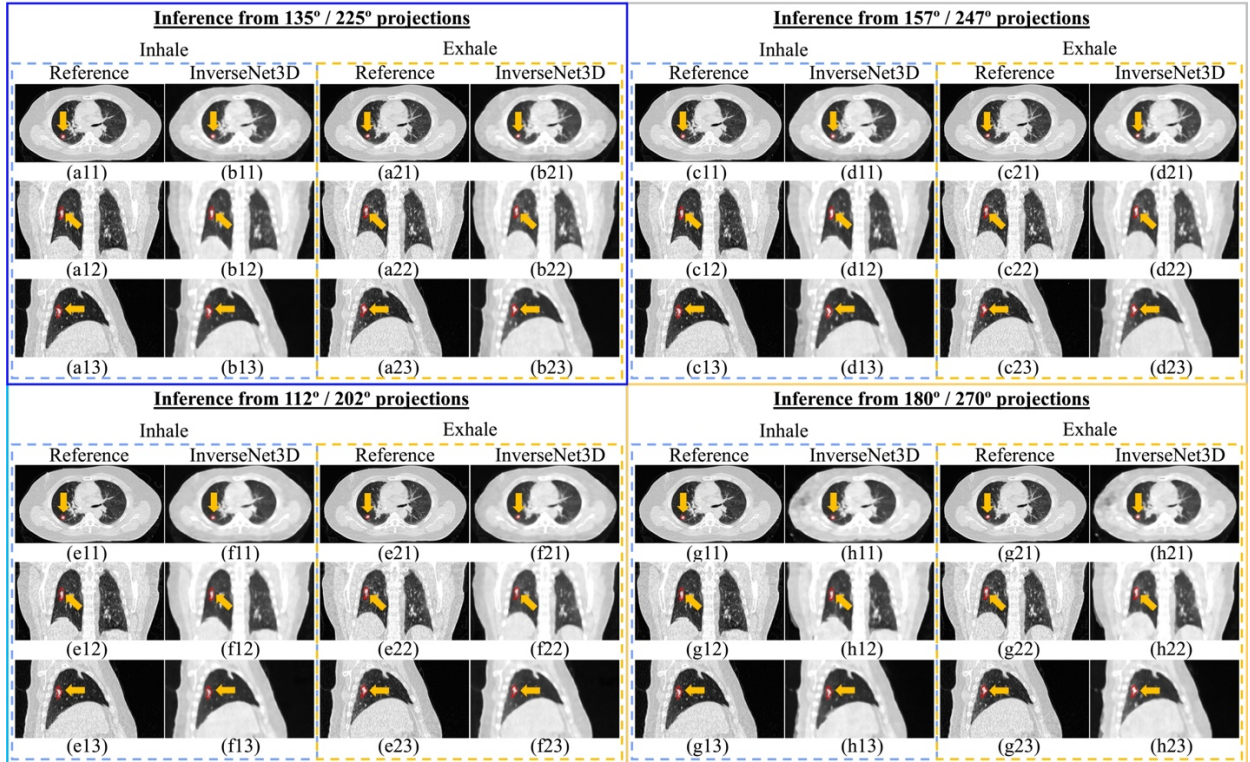

**Figure 2.** Reference and generated (InverseNet3D) volumetric images demonstrated in transversal, coronal, and sagittal views of patient 3 for inhale and exhale phases using orthogonal kV projections at various source angle pairs of 135°/225°, 157°/247°, 112°/202°, and 180°/270°. The window level of each image is [-1000, 200] Hounsfield units (HU). The orange arrows indicate the lesion ROI.

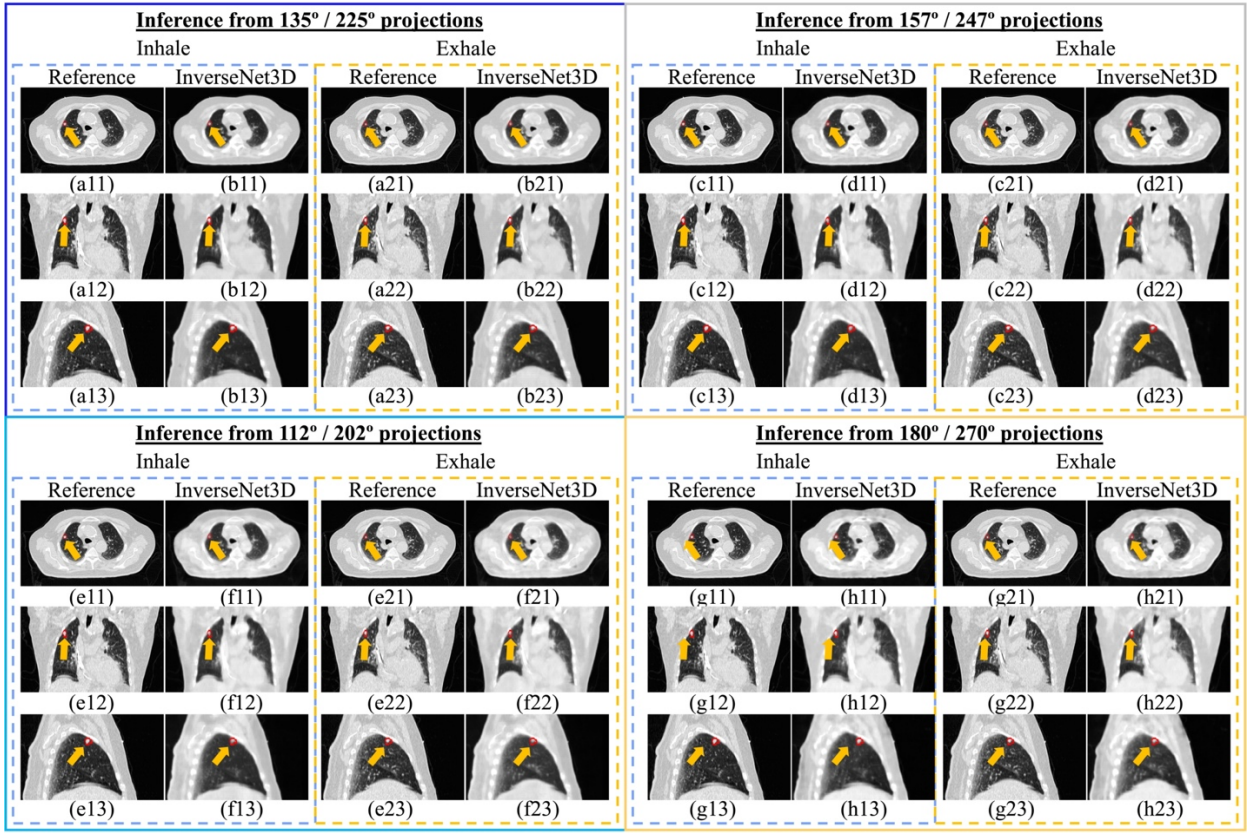

**Figure 3.** Reference and generated (InverseNet3D) volumetric images demonstrated in transversal, coronal, and sagittal views of patient 15 for inhale and exhale phases using orthogonal kV projections at various source angle pairs of 135°/225°, 157°/247°, 112°/202°, and 180°/270°. The window level of each image is [-1000, 200] Hounsfield units (HU). The orange arrows indicate the lesion ROI.

Figure 4 shows the histogram comparisons of CT numbers between the reference and InverseNet3D for patient 3 and patient 15. Figure 4(a11)/(a21)/(b11)/(b21) show that the generated images from the 135°/225° projection pair align well with the reference histograms for both inhale and exhale scenarios. The 157°/247° and 112°/202° projection pairs exhibit a slight shift from the reference histogram. Compared to other projection pairs, apparent histogram shifts of generated images from the 180°/270° projection pair can be observed in Figure 4(a14)/(a15)/(b14)/(b15) for both inhale and exhale scenarios.

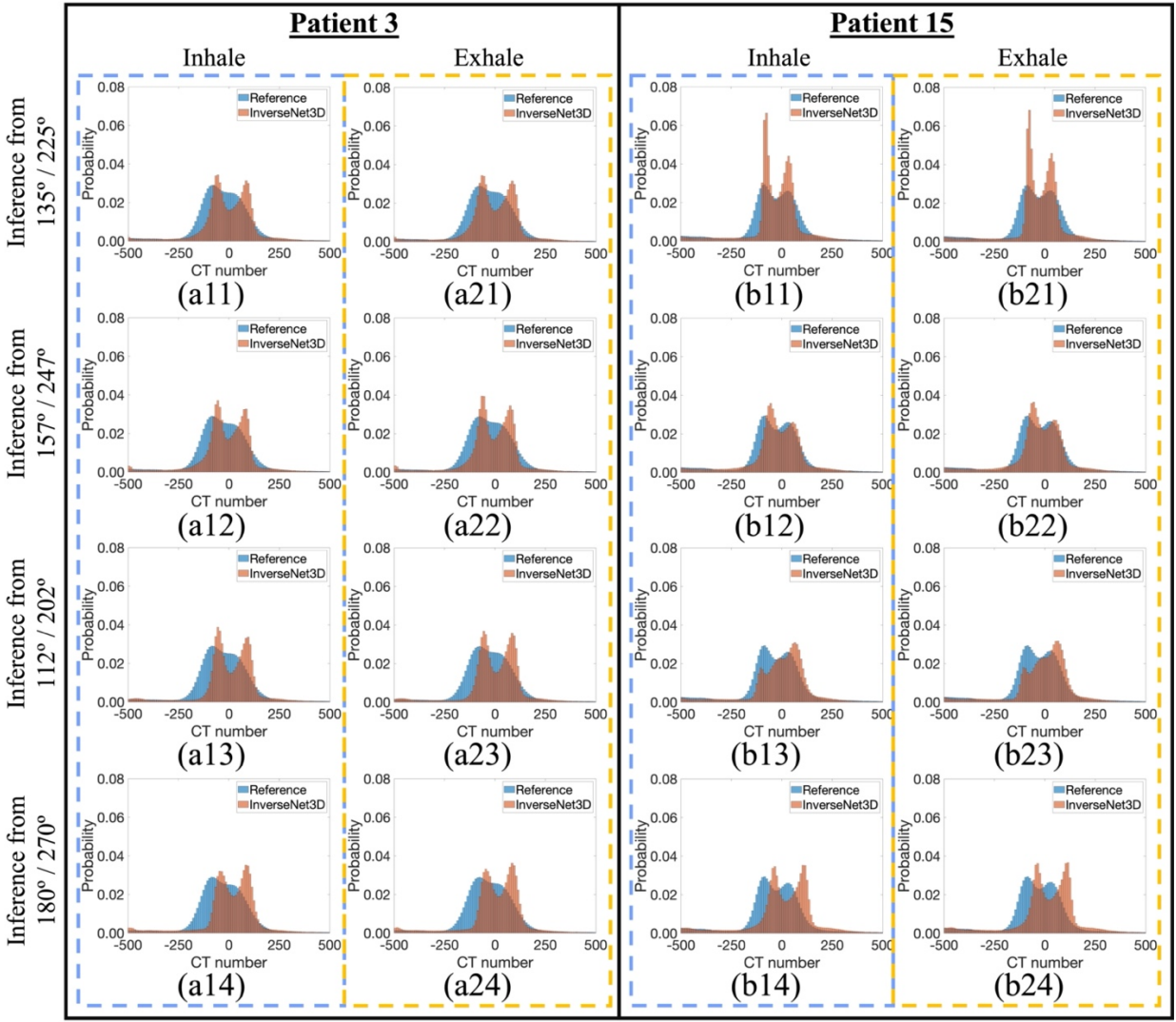

**Figure 4.** CT number distributions for inhale and exhale phases from the reference and generated (InverseNet3D) volumetric images of patient 3 and 15 using orthogonal kV projections at various source angle pairs of (a11)/(a21)/(b11)/(b21) 135°/225°, (a12)/(a22)/(b12)/(b22) 157°/247°, (a13)/(a23)/(b13)/(b23) 112°/202°, and (a14)/(a24)/(b14)/(b24) 180°/270°.

### 3.2 Treatment evaluation using water equivalent thickness (WET)

The phase-averaged WET analysis results for each patient are given in Table C1-C4 in Appendix C. Table 2 summarizes the patient-averaged WET difference ($\Delta$WET) and relative WET difference ($\varepsilon_{WET}$) results calculated by Eq. (4) and Eq. (5) for generated volumetric images by InverseNet3D using multiple orthogonal projection pairs. The volumetric images from the 180°/270° projection pair result in a minimal $\Delta$WET of -0.7 mm. However, its standard deviation is approximately twice that of images generated from the 135°/225° projection pair. The images generated from the 135°/225° projection pair also lead to the minimal standard deviations of $\Delta$WET and $\varepsilon_{WET}$ with values of 3.7 mm and 4.1%.

**Table 2.** Treatment evaluation metrics of volumetric image quality generated from InverseNet3D using the difference and relative difference of water equivalent thicknesses (ΔWET/ε$_{WET}$). All metrics are averaged over all patients given in Table C1-C4. The ΔWET and ε$_{WET}$ are calculated within the target contour for an anteroposterior proton beam.

|  | Orthogonal projections | | | |
| --- | --- | --- | --- | --- |
|  | 135° / 225° | 157° / 247° | 112° / 202° | 180° / 270° |
| ΔWET | -1.3 mm ± 3.7 mm | -1.6 mm ± 6.6 mm | -0.8 mm ± 6.3 mm | -0.7 mm ± 6.4 mm |
| ε$_{WET}$ | -1.3% ± 4.1% | -1.1% ± 5.3% | 0.3% ± 5.5% | 0.4% ± 6.8% |

Figure 5 depicts the WET distributions within the target ROI for patient 3 and patient 15 at inhale and exhale phases using InverseNet3D images with different orthogonal projection pairs. The absolute mean WET differences between inhale and exhale phases are 1 mm and 1.4 mm for patient 3 and patient 15. For patient 3, no significant variation was found for mean WET values. However, Figure 5(a12)/(a22) show that the images generated from the 157°/247° projection pair yield the maximum standard deviation of WET. The standard deviation is approximately 2 mm more extensive than the images derived from other projection pairs. For patient 15, Figure 5(b13)/(b23) display that the images generated from 180°/270° projection pair cause an approximate mean WET shift of 1.5 mm, compared to the 135°/225° projection pair images in Figure 5(b11)/(b21).

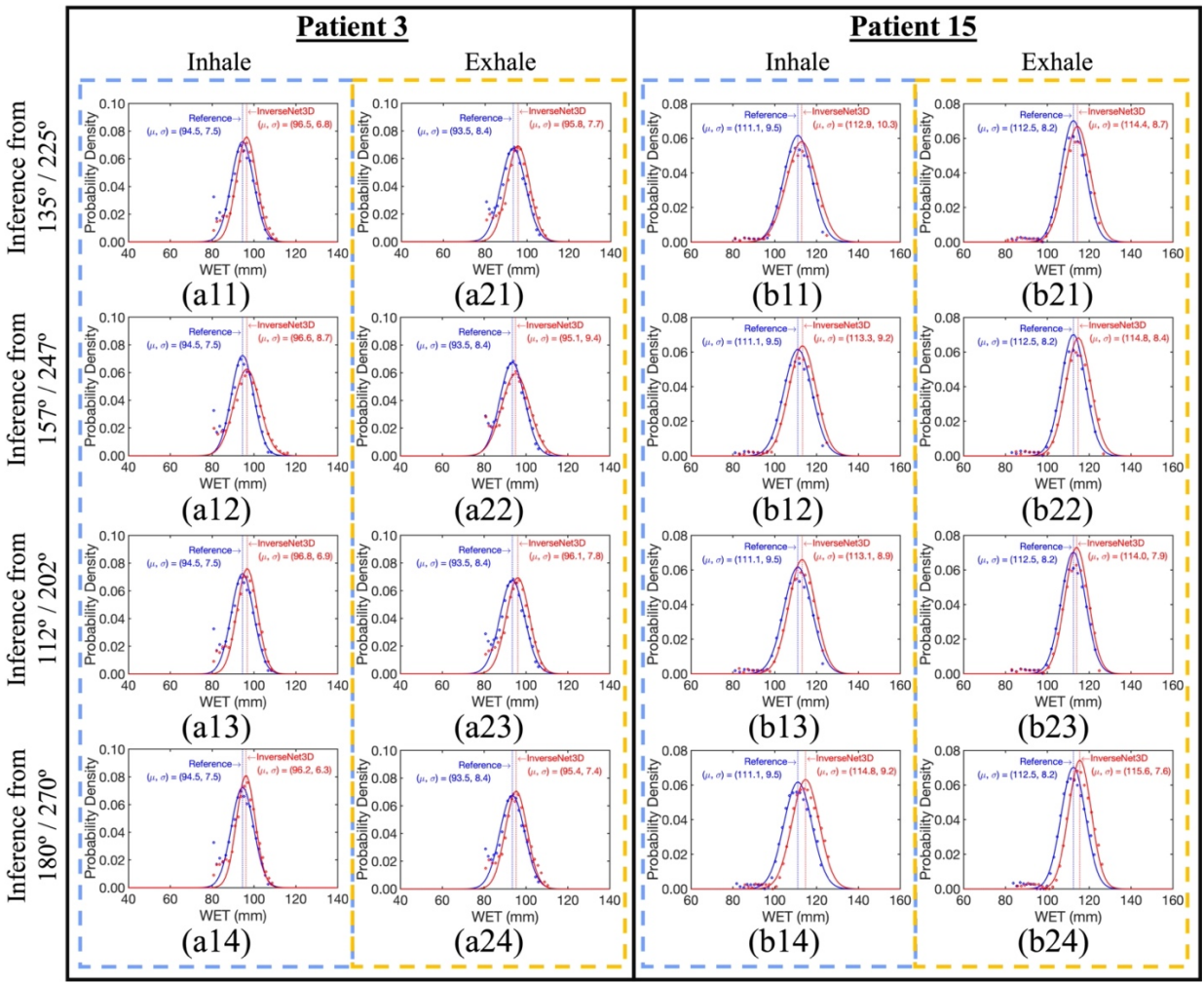

**Figure 5.** Water equivalent thickness (WET) distributions for inhale and exhale phases from the reference and generated (InverseNet3D) volumetric images of patient 3 and 15 using orthogonal kV projections at various source angle pairs of (a11)/(a21)/(b11)/(b21) 135°/225°, (a12)/(a22)/(b12)/(b22) 157°/247°, (a13)/(a23)/(b13)/(b23) 112°/202°, and (a14)/(a24)/(b14)/(b24) 180°/270°.

## 4 Discussion

Proton FLASH treatment can potentially create a paradigm shift in radiotherapy due to its capability to increase TCP while maximally reducing the toxicity for normal tissues. Besides, the inherent ultra-high dose rate feature makes the proton FLASH favorable to stereotactic body radiation therapy, which can help proton therapy become affordable and benefit patients by avoiding short-term and long-term side effects. As the decrease of fractionated treatment times, the prescription dose per fraction will inevitability increase such that an accurate image-guided system becomes essential. This work demonstrates a DL-based image-guided framework to generate fast volumetric images without motion artifacts due to the use of two instant-captured orthogonal x-ray projections. Acquiring kV x-ray projections is less than a second, and then the proposed framework can almost instantaneously deliver volumetric images for treatment evaluation.

In contrast, the current proton on-board CBCT (Stanforth *et al.*, 2022) requires approximately 35 and 60 seconds for full-fan and half-fan scans. InverseNet3D is currently implemented in the framework. Based on the retrospective patient study, all 30 patients' anatomy can be identified from the generated volumetric images, including tumor tissues. The proposed framework delivers patient-averaged MAE of approximately 75 HU, which improves the image quality by at least 25% from the previous work (Chang *et al.*, 2022d; Chang *et al.*, 2022e). Most importantly, the framework integrates a WET analysis module for the treatment evaluation, and a patient-averaged ΔWET of ~1 mm can be achieved in the present work. The analyses of WET not only indicate the image quality but also monitor the potential anatomy changes on the treatment beam path, which potentially increases the usability of the proposed framework to inform proton FLASH treatment.

Table C1-C4 provides the phase-averaged evaluation for the generated volumetric images from each patient using a leave-phase-out cross-validation method to investigate the robustness of the InverseNet3D module in the framework. Since each patient included ten raspatory phases, ten variants of InverseNet3D were trained for each patient, and a total of 300 variants of InverseNet3D were explored for a 30-patient cohort to ensure the method's robustness. Table C1-C4 indicates that the intra-patient standard deviations of each evaluation metric are usually smaller than 5% of their mean values. This result shows the proposed framework can consistently infer volumetric images from orthogonal x-ray projections. Table 1-2 provide the patient-averaged image and WET evaluation to investigate the inter-patient variability. Table 1 shows that the 180°/270° projection pair yield the most considerable inter-patient image intensity variation due to the largest MAE of 80±24 HU. The 180°/270° projection pair also results in the smallest SSIM, while the other three projection pairs have comparable SSIM values. Table 2 also shows that the 180°/270° projection pair makes the largest $\varepsilon_{WET}$ standard deviation of 6.8%. Meanwhiles, Table 2 indicates 135°/225° projection pair can derive the volumetric images with minimal inter-patient variation when considering the uncertainty. The 157°/247° and 112°/202° projection pairs show comparable results. Conclusively, these findings suggest avoiding lateral x-ray projections for image inference. The lateral projections may be less informative because the projection area is smaller compared to other projection areas from different source angles.

Figure 2-3 demonstrate the quality of volumetric image reconstruction for patients with different body sizes, target volumes, and lesion locations. The generated 3D images from the framework can successfully conserve the patient's anatomy, and both tumors and organs at risk can be identified. The results indicate

that it is feasible to use the proposed method to localize treatment regions and guide the delivery of proton beams for potential proton FLASH therapy. Another potential volumetric image application is for CT artifact reduction induced by surgical implants (Charyyev *et al.*, 2021; Chang *et al.*, 2022a; Chang *et al.*, 2022c). The kV x-ray projection provides better image quality compared to CT images when metal implants are present in patients. Reconstructing the volumetric images from kV projections can potentially mitigate the impact of photon starvation, which can cause metal artifacts on CT images. High-quality volumetric images can also increase the accuracy of material characterization to make an accurate dose evaluation (Chang *et al.*, 2022f) and support medical decision-making.

Inverse model inference from data (DL) is ill-posed, and the solution cannot satisfy the Hadamard principle of well-posedness (Koch and Lasiecka, 2002). Cross-validation is essential to ensure a robust DL model (O'Sullivan, 1986). However, this approach requires a considerable amount of numerical experiment time. Future investigation will likely focus on developing advanced validation experiments using human-mimicking phantoms and state-of-the-art instrumentation (Zhou *et al.*, 2022b) to quantify proton range uncertainty. Then the experiment data can be used to identify which DL models can work compatibly, effectively, and robustly with the proposed image-guided framework for proton FLASH radiotherapy.

## 5 Conclusions

A DL-based image-guided framework has been demonstrated for generating volumetric images using two orthogonal kV x-ray projections. The approach includes image quality and WET analyses for potential online dose evaluation and potential patient anatomy changes during inter-fractional and intra-fractional treatments. The proposed framework can inherently avoid motion artifacts and deliver instant patient anatomy evaluation to guide the treatment delivery system and inform the potential proton FLASH treatment with the target position.

**Acknowledgments**

This research is supported in part by the National Institutes of Health under Award Number R01CA215718 and R01EB032680.

**Ethical Statement**

Emory IRB review board approval was obtained, and informed consent was not required for this Health Insurance Portability and Accountability Act (HIPAA) compliant retrospective analysis.

## Appendix A. Model structure of InveseNet3D

Table A1 gives the model structures and parameters of InverseNet3D.

**Table A1.** Model form and model parameters of InverseNet3D. "Conv.", "Res.", and "Deconv." denote the convolutional, residual, and deconvolutional layers.

| | Network | Layer | Number of channels | Kernel size | Stride |
|---|---|---|---|---|---|
| Feature Extractor | Orthogonal image 1 | Conv./Conv. | 32/32 | 3/3 | 1/1 |
| | Orthogonal image 2 | Conv./Conv. | 32/32 | 3/3 | 1/1 |
| Residual Blocks | Block 1 | Conv./Conv./Res. | 32/32/32 | 4/3/4 | 2/1/2 |
| | Block 2 | Conv./Conv./Res. | 32/32/32 | 4/3/4 | 2/1/2 |
| | Block 3 | Conv./Conv./Res. | 32/32/32 | 4/3/4 | 2/1/2 |
| | Block 4 | Conv./Conv./Res. | 32/32/32 | 4/3/4 | 2/1/2 |
| | Block 5 | Conv./Conv./Res. | 32/32/32 | 4/3/4 | 2/1/2 |
| Deconvolutional Blocks | Deconvolution | Deconv./Deconv./Deconv./Deconv. | 32/64/128/256 | 3/3/3/3 | 2/2/2/2 |

## Appendix B. Hounsfield look-up table.

Table B1 shows the Hounsfield look-up table to convert CT numbers to material relative stopping power ratios.

**Table B1**. Hounsfield look-up table for relative stopping power.

| CT number (Hounsfield unit) | Relative stopping power |
|---|---|
| -1024 | $9.0 \times 10^{-4}$ |
| -980 | $1.0 \times 10^{-3}$ |
| -741 | 0.257 |
| -707 | 0.288 |
| -560 | 0.432 |
| -93 | 0.891 |
| -61 | 0.916 |
| -48 | 0.929 |
| -24 | 0.958 |
| 0 | 1.000 |
| 19 | 1.001 |
| 29 | 1.002 |
| 48 | 1.045 |
| 52 | 1.049 |
| 76 | 1.067 |
| 101 | 1.097 |
| 189 | 1.094 |
| 200 | 1.103 |
| 242 | 1.149 |
| 383 | 1.263 |
| 427 | 1.289 |
| 549 | 1.384 |
| 565 | 1.394 |
| 628 | 1.430 |
| 702 | 1.495 |
| 761 | 1.513 |
| 829 | 1.586 |
| 923 | 1.656 |
| 1157 | 1.780 |
| 1260 | 1.898 |
| 2495 | 2.685 |

# Appendix C. Model structure of InveseNet3D

Table C1-C4 gives the metrics for image quality and treatment evaluations, which are averaged over all phases for each patient.

**Table C1**. Evaluation metrics of volumetric image quality generated from InverseNet3D using orthogonal images with the source angle pair of 135°-225°. The metrics include mean error (ME), mean absolute error (MAE), peak signal-to-noise ratio (PSNR), and structural similarity index measure (SSIM). All metrics are averaged over all phases for each patient. ME and MAE are evaluated for the whole volume, and PSNR and SSIM are computed for the target contour. The difference and relative difference of water equivalent thicknesses ($\Delta$WET/$\varepsilon_{WET}$) are calculated within the target contour for an anteroposterior proton beam.

| Patient | ME (HU) | MAE (HU) | PSNR (dB) | SSIM | $\Delta$WET (mm) | $\varepsilon_{WET}$ |
|---|---|---|---|---|---|---|
| 1 | -8.5 ± 0.2 | 42.3 ± 0.8 | 22.3 ± 0.6 | 0.967 ± 0.011 | -2.6 ± 0.2 | -3.7% ± 0.2% |
| 2 | 11.6 ± 0.5 | 50.2 ± 0.6 | 20.5 ± 0.5 | 0.978 ± 0.002 | -2.8 ± 0.2 | -2.0% ± 0.2% |
| 3 | 34.2 ± 0.4 | 59.3 ± 0.3 | 14.7 ± 0.2 | 0.979 ± 0.005 | 1.3 ± 0.2 | 1.8% ± 0.3% |
| 4 | -37.8 ± 2.9 | 50.3 ± 3.7 | 12.2 ± 0.7 | 0.961 ± 0.012 | -2.2 ± 0.2 | -4.4% ± 0.4% |
| 5 | 0.5 ± 0.3 | 53.0 ± 0.1 | 13.7 ± 0.1 | 0.989 ± 0.003 | 4.1 ± 0.1 | 5.2% ± 0.1% |
| 6 | -23.7 ± 0.1 | 45.2 ± 0.2 | 21.1 ± 0.3 | 0.960 ± 0.007 | -2.2 ± 0.1 | -5.6% ± 0.2% |
| 7 | -13.4 ± 0.7 | 76.9 ± 3.1 | 17.2 ± 0.5 | 0.967 ± 0.010 | -1.2 ± 0.3 | -3.1% ± 0.7% |
| 8 | -22.3 ± 1.1 | 96.1 ± 4.1 | 16.9 ± 0.4 | 0.944 ± 0.009 | -2.3 ± 0.7 | -2.6% ± 0.8% |
| 9 | -5.7 ± 0.3 | 88.5 ± 7.1 | 14.9 ± 0.4 | 0.892 ± 0.016 | 2.7 ± 0.9 | 2.2% ± 0.8% |
| 10 | -4.2 ± 1.0 | 99.1 ± 0.6 | 19.0 ± 0.2 | 0.901 ± 0.004 | -7.1 ± 0.8 | -7.4% ± 0.5% |
| 11 | -11.0 ± 0.5 | 102.8 ± 1.3 | 15.8 ± 0.1 | 0.866 ± 0.005 | 0.1 ± 0.3 | 0.1% ± 0.3% |
| 12 | -3.0 ± 0.4 | 84.7 ± 2.2 | 15.4 ± 0.4 | 0.900 ± 0.006 | -5.8 ± 0.4 | -7.6% ± 0.5% |
| 13 | -4.7 ± 0.6 | 69.9 ± 9.3 | 17.7 ± 1.4 | 0.988 ± 0.007 | 1.3 ± 1.0 | 4.5% ± 3.5% |
| 14 | -5.8 ± 0.5 | 84.0 ± 0.4 | 18.6 ± 0.1 | 0.882 ± 0.007 | 0.7 ± 0.3 | 2.1% ± 0.9% |
| 15 | -5.3 ± 0.3 | 38.5 ± 0.1 | 22.3 ± 0.2 | 0.979 ± 0.005 | 1.3 ± 0.3 | 1.3% ± 0.3% |
| 16 | -73.0 ± 0.4 | 94.5 ± 0.9 | 19.0 ± 0.2 | 0.992 ± 0.009 | -0.9 ± 0.2 | -3.3% ± 0.5% |
| 17 | 14.4 ± 0.3 | 55.7 ± 0.7 | 21.1 ± 0.5 | 0.899 ± 0.004 | -4.8 ± 0.5 | -5.6% ± 0.5% |
| 18 | -42.8 ± 4.2 | 57.0 ± 5.8 | 18.8 ± 0.4 | 0.923 ± 0.005 | -3.1 ± 1.0 | -2.8% ± 0.9% |
| 19 | -6.1 ± 0.3 | 47.1 ± 0.8 | 21.0 ± 0.4 | 0.951 ± 0.003 | -2.5 ± 0.1 | -5.6% ± 0.1% |
| 20 | -21.0 ± 0.4 | 48.5 ± 0.3 | 32.8 ± 0.1 | 0.991 ± 0.002 | -0.4 ± 0.1 | -0.7% ± 0.1% |
| 21 | -28.0 ± 0.8 | 71.2 ± 0.8 | 21.4 ± 0.3 | 0.939 ± 0.012 | -1.5 ± 1.5 | -2.9% ± 2.8% |
| 22 | -30.0 ± 0.9 | 100.9 ± 1.0 | 19.2 ± 0.2 | 0.993 ± 0.001 | 0.1 ± 0.1 | 0.3% ± 0.3% |
| 23 | -17.7 ± 0.8 | 80.0 ± 0.9 | 20.1 ± 0.1 | 0.979 ± 0.001 | -11.6 ± 1.3 | -7.0% ± 0.7% |
| 24 | -22.0 ± 1.2 | 96.0 ± 1.5 | 21.4 ± 0.1 | 0.911 ± 0.010 | 1.1 ± 0.7 | 1.0% ± 0.6% |
| 25 | -11.9 ± 0.3 | 76.0 ± 1.3 | 15.4 ± 0.1 | 0.828 ± 0.074 | 5.1 ± 0.7 | 4.4% ± 0.6% |
| 26 | -29.4 ± 0.6 | 103.8 ± 0.7 | 18.6 ± 0.2 | 0.901 ± 0.009 | -1.7 ± 1.1 | -2.2% ± 1.4% |
| 27 | -22.3 ± 1.3 | 87.5 ± 3.1 | 20.0 ± 0.7 | 0.945 ± 0.004 | -1.0 ± 0.7 | -2.4% ± 1.6% |
| 28 | -32.4 ± 0.5 | 103.8 ± 0.3 | 21.6 ± 0.04 | 0.945 ± 0.005 | 2.1 ± 0.4 | 5.3% ± 1.0% |
| 29 | -20.0 ± 0.4 | 105.5 ± 1.9 | 18.4 ± 0.3 | 0.892 ± 0.004 | 3.5 ± 0.9 | 7.3% ± 1.9% |
| 30 | -23.9 ± 1.4 | 100.0 ± 3.9 | 18.7 ± 0.6 | 0.898 ± 0.017 | -9.2 ± 2.0 | -4.8% ± 1.0% |

16arXiv: 2210.00971

**Table C2**. Evaluation metrics of volumetric image quality generated from InverseNet3D using orthogonal images with the source angle pair of 157°-247°. The metrics include mean error (ME), mean absolute error (MAE), peak signal-to-noise ratio (PSNR), and structural similarity index measure (SSIM). All metrics are averaged over all phases for each patient. ME and MAE are evaluated for the whole volume, and PSNR and SSIM are computed for the target contour. The difference and relative difference of water equivalent thicknesses ($\Delta$WET/$\varepsilon_{WET}$) are calculated within the target contour for an anteroposterior proton beam.

| Patient | ME (HU) | MAE (HU) | PSNR (dB) | SSIM | $\Delta$WET (mm) | $\varepsilon_{WET}$ |
|---|---|---|---|---|---|---|
| 1 | -11.7 ± 0.7 | 39.2 ± 0.5 | 20.7 ± 0.8 | 0.962 ± 0.014 | -2.7 ± 0.3 | -3.8% ± 0.3% |
| 2 | 9.8 ± 0.8 | 48.2 ± 0.3 | 19.1 ± 0.5 | 0.977 ± 0.003 | -3.1 ± 0.3 | -2.2% ± 0.2% |
| 3 | 30.5 ± 1.1 | 56.3 ± 1.0 | 16.3 ± 0.3 | 0.968 ± 0.011 | 1.8 ± 0.7 | 2.5% ± 0.9% |
| 4 | -34.9 ± 3.5 | 52.3 ± 5.3 | 15.7 ± 0.8 | 0.964 ± 0.013 | -2.1 ± 0.2 | -4.2% ± 0.4% |
| 5 | -5.1 ± 0.1 | 55.2 ± 0.1 | 17.7 ± 0.1 | 0.986 ± 0.001 | 2.4 ± 0.2 | 3.1% ± 0.3% |
| 6 | -26.2 ± 0.4 | 40.4 ± 0.4 | 21.6 ± 0.2 | 0.965 ± 0.008 | -2.1 ± 0.1 | -5.3% ± 0.2% |
| 7 | -13.2 ± 0.8 | 83.5 ± 2.4 | 15.8 ± 0.4 | 0.795 ± 0.011 | 3.9 ± 0.6 | 10.2% ± 1.6% |
| 8 | -15.5 ± 1.8 | 100.5 ± 3.6 | 16.2 ± 0.4 | 0.966 ± 0.006 | -1.4 ± 0.8 | -1.5% ± 1.0% |
| 9 | -4.4 ± 0.4 | 91.4 ± 5.6 | 14.1 ± 0.4 | 0.877 ± 0.012 | 6.6 ± 1.2 | 5.4% ± 1.0% |
| 10 | 6.4 ± 0.4 | 95.9 ± 1.3 | 15.7 ± 0.7 | 0.942 ± 0.007 | -4.1 ± 1.2 | -4.3% ± 1.4% |
| 11 | -10.1 ± 0.4 | 97.2 ± 1.4 | 15.7 ± 0.0 | 0.966 ± 0.004 | 0.4 ± 0.5 | 0.3% ± 0.5% |
| 12 | -1.8 ± 0.3 | 82.4 ± 2.8 | 16.7 ± 0.5 | 0.956 ± 0.004 | 0.4 ± 0.9 | 0.5% ± 1.1% |
| 13 | -7.3 ± 0.8 | 69.7 ± 7.4 | 17.4 ± 1.3 | 0.985 ± 0.014 | 0.7 ± 0.9 | 2.4% ± 3.3% |
| 14 | -4.7 ± 0.2 | 87.5 ± 0.5 | 17.6 ± 0.4 | 0.855 ± 0.011 | 0.7 ± 0.2 | 2.1% ± 0.7% |
| 15 | 12.2 ± 0.1 | 42.6 ± 0.2 | 29.4 ± 0.2 | 0.986 ± 0.003 | 2.1 ± 0.2 | 2.1% ± 0.2% |
| 16 | -53.4 ± 0.3 | 75.1 ± 0.3 | 19.3 ± 0.1 | 0.992 ± 0.007 | -0.8 ± 0.2 | -3.2% ± 0.4% |
| 17 | 10.8 ± 0.1 | 63.1 ± 0.5 | 22.5 ± 0.5 | 0.956 ± 0.006 | -3.8 ± 0.6 | -4.4% ± 0.5% |
| 18 | -24.9 ± 2.0 | 52.3 ± 3.6 | 24.1 ± 4.8 | 0.955 ± 0.005 | -3.0 ± 0.5 | -2.6% ± 0.5% |
| 19 | -7.5 ± 0.2 | 49.6 ± 0.5 | 28.5 ± 0.4 | 0.962 ± 0.003 | -2.4 ± 0.0 | -5.4% ± 0.1% |
| 20 | 1.8 ± 0.2 | 40.6 ± 0.2 | 32.4 ± 0.1 | 0.992 ± 0.001 | 0.6 ± 0.1 | 1.3% ± 0.1% |
| 21 | -13.4 ± 0.7 | 80.7 ± 1.1 | 20.6 ± 0.3 | 0.945 ± 0.011 | -1.5 ± 0.8 | -3.2% ± 1.6% |
| 22 | -15.1 ± 0.8 | 95.7 ± 1.0 | 23.0 ± 0.2 | 0.987 ± 0.001 | -0.2 ± 0.1 | -0.6% ± 0.5% |
| 23 | -7.9 ± 0.2 | 92.6 ± 1.3 | 23.4 ± 2.7 | 0.961 ± 0.002 | -25.9 ± 1.0 | -15.7% ± 0.4% |
| 24 | -10.5 ± 0.7 | 80.6 ± 1.2 | 21.5 ± 0.1 | 0.966 ± 0.008 | 1.7 ± 0.8 | 1.6% ± 0.7% |
| 25 | -13.7 ± 0.6 | 82.0 ± 0.7 | 22.5 ± 5.9 | 0.872 ± 0.009 | 7.4 ± 0.4 | 6.4% ± 0.3% |
| 26 | -10.4 ± 0.3 | 91.0 ± 0.6 | 20.9 ± 0.05 | 0.903 ± 0.015 | -3.0 ± 0.8 | -4.1% ± 0.9% |
| 27 | -13.7 ± 1.0 | 80.2 ± 1.6 | 18.2 ± 0.7 | 0.934 ± 0.003 | -3.0 ± 0.6 | -7.0% ± 1.3% |
| 28 | -22.3 ± 0.4 | 93.0 ± 0.2 | 22.0 ± 0.9 | 0.963 ± 0.003 | 2.1 ± 0.2 | 5.3% ± 0.5% |
| 29 | -8.7 ± 1.2 | 109.1 ± 1.8 | 18.1 ± 0.3 | 0.892 ± 0.006 | 1.2 ± 0.4 | 2.5% ± 0.9% |
| 30 | -12.3 ± 0.7 | 93.9 ± 2.6 | 18.7 ± 0.6 | 0.928 ± 0.007 | -20.9 ± 1.9 | -10.9% ± 1.0% |

**Table C3.** Evaluation metrics of volumetric image quality generated from InverseNet3D using orthogonal images with the source angle pair of 112°-202°. The metrics include mean error (ME), mean absolute error (MAE), peak signal-to-noise ratio (PSNR), and structural similarity index measure (SSIM). All metrics are averaged over all phases for each patient. ME and MAE are evaluated for the whole volume, and PSNR and SSIM are computed for the target contour. The difference and relative difference of water equivalent thicknesses ($\Delta$WET/ $\varepsilon_{WET}$) are calculated within the target contour for an anteroposterior proton beam.

| Patient | ME (HU) | MAE (HU) | PSNR (dB) | SSIM | $\Delta$WET (mm) | $\varepsilon_{WET}$ |
|---|---|---|---|---|---|---|
| 1 | 1.6 ± 0.3 | 45.2 ± 0.3 | 18.4 ± 0.3 | 0.960 ± 0.017 | -3.0 ± 0.8 | -4.1% ± 0.9% |
| 2 | 19.7 ± 0.5 | 56.0 ± 0.2 | 18.1 ± 0.8 | 0.977 ± 0.004 | -2.9 ± 0.3 | -2.0% ± 0.2% |
| 3 | 43.4 ± 0.6 | 64.3 ± 0.2 | 15.5 ± 0.4 | 0.973 ± 0.007 | 0.9 ± 0.2 | 1.2% ± 0.3% |
| 4 | -23.5 ± 2.9 | 43.6 ± 3.4 | 18.4 ± 0.8 | 0.970 ± 0.010 | -1.8 ± 0.1 | -3.7% ± 0.2% |
| 5 | 9.3 ± 0.2 | 49.7 ± 0.1 | 26.3 ± 0.6 | 0.991 ± 0.002 | 3.2 ± 0.2 | 4.1% ± 0.2% |
| 6 | -19.0 ± 0.1 | 44.8 ± 0.1 | 18.5 ± 0.1 | 0.952 ± 0.008 | -1.7 ± 0.2 | -4.3% ± 0.2% |
| 7 | -2.9 ± 0.2 | 72.0 ± 2.0 | 13.0 ± 0.4 | 0.940 ± 0.010 | 0.4 ± 0.5 | 1.1% ± 1.4% |
| 8 | -7.0 ± 0.5 | 92.7 ± 3.7 | 15.5 ± 0.4 | 0.945 ± 0.006 | -0.3 ± 0.9 | -0.4% ± 1.0% |
| 9 | 8.5 ± 0.9 | 91.7 ± 6.8 | 8.6 ± 0.2 | 0.908 ± 0.009 | -5.7 ± 1.4 | -4.6% ± 1.2% |
| 10 | 13.4 ± 0.3 | 100.1 ± 2.0 | 16.6 ± 0.2 | 0.919 ± 0.002 | -6.5 ± 0.3 | -6.8% ± 0.2% |
| 11 | 0.3 ± 0.3 | 92.0 ± 1.4 | 9.3 ± 0.1 | 0.942 ± 0.006 | 2.8 ± 0.4 | 2.4% ± 0.4% |
| 12 | 12.0 ± 0.6 | 86.2 ± 2.7 | 10.7 ± 0.3 | 0.915 ± 0.004 | -2.4 ± 0.4 | -3.1% ± 0.5% |
| 13 | 7.9 ± 2.3 | 67.8 ± 5.5 | 12.7 ± 0.8 | 0.991 ± 0.004 | 0.7 ± 0.9 | 2.6% ± 3.4% |
| 14 | 1.9 ± 0.2 | 86.3 ± 0.5 | 12.2 ± 0.1 | 0.886 ± 0.011 | 0.6 ± 0.3 | 1.8% ± 0.9% |
| 15 | 23.4 ± 0.4 | 48.5 ± 0.4 | 16.2 ± 0.1 | 0.979 ± 0.004 | 1.3 ± 0.3 | 1.3% ± 0.3% |
| 16 | -50.3 ± 0.4 | 74.0 ± 0.6 | 15.6 ± 0.3 | 0.996 ± 0.006 | 0.4 ± 0.1 | 1.3% ± 0.3% |
| 17 | 32.7 ± 0.4 | 76.0 ± 0.4 | 17.7 ± 0.6 | 0.885 ± 0.011 | 11.2 ± 3.6 | 12.8% ± 3.5% |
| 18 | -18.3 ± 2.5 | 44.3 ± 5.8 | 15.7 ± 0.5 | 0.957 ± 0.003 | -0.3 ± 1.0 | -0.3% ± 0.9% |
| 19 | 11.3 ± 0.3 | 56.3 ± 1.2 | 14.6 ± 0.3 | 0.961 ± 0.003 | -1.8 ± 0.1 | -4.1% ± 0.2% |
| 20 | -0.4 ± 0.6 | 44.6 ± 0.3 | 17.7 ± 0.1 | 0.983 ± 0.002 | -1.0 ± 0.1 | -1.9% ± 0.2% |
| 21 | 1.3 ± 0.6 | 74.7 ± 1.1 | 19.0 ± 0.6 | 0.941 ± 0.015 | 1.7 ± 1.7 | 3.8% ± 3.8% |
| 22 | -8.4 ± 0.6 | 84.8 ± 1.0 | 16.7 ± 0.2 | 0.969 ± 0.003 | 0.8 ± 0.1 | 2.9% ± 0.3% |
| 23 | 12.4 ± 0.2 | 84.2 ± 1.0 | 16.8 ± 0.1 | 0.968 ± 0.002 | -13.7 ± 1.1 | -8.3% ± 0.7% |
| 24 | 23.0 ± 0.7 | 88.9 ± 1.4 | 15.6 ± 0.1 | 0.947 ± 0.007 | 5.1 ± 1.7 | 5.0% ± 1.8% |
| 25 | 12.1 ± 0.9 | 92.9 ± 1.7 | 9.4 ± 0.1 | 0.775 ± 0.021 | 1.8 ± 1.5 | 1.5% ± 1.3% |
| 26 | 1.2 ± 0.5 | 96.2 ± 0.9 | 16.9 ± 0.1 | 0.915 ± 0.007 | 3.7 ± 0.4 | 5.0% ± 0.6% |
| 27 | -5.2 ± 0.3 | 71.0 ± 1.6 | 15.9 ± 0.5 | 0.934 ± 0.004 | -0.9 ± 0.8 | -2.2% ± 1.9% |
| 28 | 0.5 ± 0.3 | 89.8 ± 0.3 | 19.1 ± 0.1 | 0.953 ± 0.004 | 5.0 ± 0.2 | 12.4% ± 0.6% |
| 29 | -1.4 ± 0.7 | 82.2 ± 1.5 | 18.8 ± 0.3 | 0.886 ± 0.004 | 3.9 ± 0.6 | 8.1% ± 1.2% |
| 30 | 16.2 ± 1.0 | 100.5 ± 3.3 | 17.7 ± 0.7 | 0.905 ± 0.006 | -25.0 ± 2.5 | -13.0% ± 1.3% |

**Table C4**. Evaluation metrics of volumetric image quality generated from InverseNet3D using orthogonal images with the source angle pair of 180°-270°. The metrics include mean error (ME), mean absolute error (MAE), peak signal-to-noise ratio (PSNR), and structural similarity index measure (SSIM). All metrics are averaged over all phases for each patient. ME and MAE are evaluated for the whole volume, and PSNR and SSIM are computed for the target contour. The difference and relative difference of water equivalent thicknesses ($\Delta$WET/ $\varepsilon_{WET}$) are calculated within the target contour for an anteroposterior proton beam.

| Patient | ME (HU) | MAE (HU) | PSNR (dB) | SSIM | $\Delta$WET (mm) | $\varepsilon_{WET}$ |
|---|---|---|---|---|---|---|
| 1 | -6.8 ± 0.4 | 33.4 ± 0.4 | 23.7 ± 1.3 | 0.966 ± 0.008 | -1.6 ± 0.4 | -2.2% ± 0.4% |
| 2 | 16.9 ± 0.8 | 44.7 ± 0.2 | 21.3 ± 0.4 | 0.965 ± 0.003 | 1.6 ± 0.4 | 1.1% ± 0.3% |
| 3 | 34.9 ± 0.8 | 60.0 ± 0.4 | 17.8 ± 0.3 | 0.970 ± 0.005 | 0.9 ± 0.5 | 1.3% ± 0.6% |
| 4 | -38.4 ± 3.7 | 60.8 ± 5.6 | 22.2 ± 0.9 | 0.967 ± 0.010 | -1.6 ± 0.3 | -3.3% ± 0.7% |
| 5 | 20.9 ± 0.4 | 100.2 ± 1.1 | 19.6 ± 0.5 | 0.954 ± 0.004 | 9.9 ± 0.3 | 12.5% ± 0.6% |
| 6 | -13.4 ± 0.3 | 35.5 ± 0.2 | 22.5 ± 0.2 | 0.948 ± 0.005 | -0.5 ± 0.1 | -1.3% ± 0.4% |
| 7 | -16.5 ± 0.8 | 78.1 ± 2.7 | 18.2 ± 0.5 | 0.708 ± 0.021 | 3.6 ± 0.7 | 9.5% ± 1.7% |
| 8 | -14.0 ± 0.8 | 98.5 ± 4.5 | 17.6 ± 0.4 | 0.913 ± 0.014 | -4.9 ± 0.9 | -5.6% ± 1.1% |
| 9 | -16.5 ± 0.9 | 90.7 ± 6.1 | 16.1 ± 0.5 | 0.884 ± 0.011 | 5.1 ± 1.3 | 4.2% ± 1.1% |
| 10 | -2.6 ± 0.3 | 93.7 ± 1.7 | 20.5 ± 0.3 | 0.920 ± 0.002 | -3.7 ± 0.5 | -3.8% ± 0.4% |
| 11 | -1.9 ± 0.3 | 96.0 ± 1.5 | 17.3 ± 0.1 | 0.839 ± 0.004 | -2.9 ± 0.3 | -2.5% ± 0.3% |
| 12 | 5.4 ± 0.6 | 80.6 ± 2.8 | 18.5 ± 0.5 | 0.950 ± 0.002 | -2.1 ± 0.5 | -2.8% ± 0.7% |
| 13 | 3.6 ± 1.7 | 65.6 ± 6.1 | 19.2 ± 1.8 | 0.957 ± 0.030 | 1.5 ± 1.0 | 5.2% ± 3.6% |
| 14 | -12.0 ± 0.3 | 84.4 ± 0.4 | 19.8 ± 0.1 | 0.872 ± 0.007 | 2.1 ± 0.2 | 6.1% ± 0.5% |
| 15 | 17.6 ± 0.4 | 45.5 ± 0.2 | 28.6 ± 0.1 | 0.982 ± 0.005 | 3.4 ± 0.2 | 3.4% ± 0.2% |
| 16 | -28.0 ± 0.4 | 82.1 ± 0.5 | 17.9 ± 0.1 | 0.996 ± 0.004 | -0.3 ± 0.1 | -1.0% ± 0.3% |
| 17 | 17.4 ± 0.7 | 62.9 ± 0.3 | 25.0 ± 0.4 | 0.942 ± 0.013 | -6.7 ± 0.7 | -7.8% ± 1.1% |
| 18 | -22.8 ± 3.2 | 44.3 ± 4.0 | 18.7 ± 0.5 | 0.971 ± 0.002 | -0.5 ± 0.7 | -0.4% ± 0.6% |
| 19 | -5.2 ± 0.6 | 55.3 ± 0.2 | 27.7 ± 0.3 | 0.960 ± 0.005 | -2.6 ± 0.1 | -5.9% ± 0.1% |
| 20 | -5.1 ± 0.4 | 120.5 ± 0.6 | 20.2 ± 0.1 | 0.917 ± 0.004 | 7.2 ± 0.4 | 14.3% ± 0.7% |
| 21 | -2.6 ± 1.2 | 86.2 ± 1.1 | 19.7 ± 0.2 | 0.949 ± 0.006 | -0.1 ± 1.2 | -0.1% ± 2.4% |
| 22 | -11.4 ± 0.9 | 104.8 ± 0.7 | 18.1 ± 0.1 | 0.884 ± 0.005 | -2.5 ± 0.1 | -8.7% ± 0.3% |
| 23 | -3.4 ± 0.2 | 92.3 ± 0.9 | 19.3 ± 0.2 | 0.969 ± 0.002 | -24.1 ± 1.3 | -14.6% ± 0.5% |
| 24 | 1.8 ± 1.4 | 93.4 ± 1.1 | 21.1 ± 0.1 | 0.968 ± 0.004 | 8.5 ± 2.0 | 8.2% ± 1.8% |
| 25 | -4.8 ± 0.3 | 85.7 ± 0.5 | 24.6 ± 2.9 | 0.845 ± 0.013 | -5.2 ± 1.1 | -4.5% ± 0.9% |
| 26 | -10.6 ± 0.7 | 98.0 ± 0.5 | 20.7 ± 0.04 | 0.928 ± 0.005 | -0.8 ± 0.5 | -1.0% ± 0.6% |
| 27 | -1.1 ± 0.7 | 92.5 ± 1.7 | 19.2 ± 0.6 | 0.959 ± 0.002 | -0.5 ± 0.6 | -1.3% ± 1.4% |
| 28 | -0.1 ± 0.7 | 99.8 ± 0.1 | 20.4 ± 0.04 | 0.966 ± 0.004 | 2.4 ± 0.5 | 5.9% ± 1.2% |
| 29 | -5.6 ± 0.5 | 118.1 ± 0.5 | 18.4 ± 0.4 | 0.875 ± 0.003 | 6.3 ± 0.6 | 13.1% ± 1.1% |
| 30 | 4.2 ± 0.8 | 106.0 ± 3.0 | 17.5 ± 0.5 | 0.929 ± 0.007 | -11.9 ± 4.4 | -6.2% ± 2.3% |